\documentclass[twocolumn]{aastex63} 
\usepackage[normalem]{ulem}
\usepackage[version=4]{mhchem}
\graphicspath{{./}{figures/}}

\shorttitle{iSLAT}
\shortauthors{Jellison et al.}

\begin{document}

\title{iSLAT: the Interactive Spectral-Line Analysis Tool for JWST and beyond}

\correspondingauthor{Andrea Banzatti}
\email{banzatti@txstate.edu}

\author[0000-0003-1582-0530]{Evan G. Jellison}
\author{Matthew Johnson}
\author[0000-0003-4335-0900]{Andrea Banzatti}
\affiliation{Department of Physics, Texas State University, 749 N Comanche Street, San Marcos, TX 78666, USA}

\author{Simon Bruderer}
\affiliation{Max-Planck-Institut für extraterrestrische Physik, Gießenbachstraße 1, 85748 Garching bei München}


\begin{abstract}
We present iSLAT (the Interactive Spectral-Line Analysis Tool), a python-based graphical tool that allows users to interactively explore and manually fit line emission observed in molecular spectra. iSLAT adopts a simple slab model that simulates emission spectra with a small set of parameters (temperature, emitting area, column density, and line broadening) that users can adjust in real time for multiple molecules or multiple thermal components of a same molecule. A central feature of iSLAT is the possibility to interactively inspect individual lines or line clusters to visualize their properties at high resolution and identify them in the population diagram. iSLAT provides a number of additional features, including the option to identify lines that are not blended at the instrumental resolution, the possibility to save custom line lists selected by the user, and to fit and measure their properties (line flux, width, and centroid) for later analysis. In this paper we launch the tool and demonstrate it on infrared spectra from the James Webb Space Telescope and ground-based instruments that provide higher resolving power. We also share curated line lists that are useful for the analysis of the forest of water emission lines observed from protoplanetary disks. iSLAT is shared with the community on GitHub. 
\end{abstract}

\section{Introduction} 
The analysis of molecular emission spectra in astrophysics can be a daunting challenge due to the overlap of spectral forests of lines, more so if multiple chemical species are present. This is especially true when spectrographs do not have the necessary resolving power to separate individual lines from each other. Famous examples of that are the molecular forests observed at infrared wavelengths with the Spitzer Space Telescope at the resolution of $R \approx 700$ with IRS \citep{irs}, where hundreds of emission blends from \ce{H2O}, \ce{OH}, \ce{HCN}, \ce{C2H2}, and \ce{CO2} were observed from the warm inner regions of protoplanetary disks \citep{cn08,cn11,pont10,salyk11_spitz}. Even at the $\times30-100$ higher resolving power provided by ground-based spectrographs, infrared molecular spectra can be dominated by line blending from different molecules and/or emission components \citep[e.g.][]{najita03,mandell12,brown13,banz22,banz23}. With the recent and upcoming improvements in resolution and spectral coverage of infrared spectrographs in space and on the ground, the analysis and interpretation of hundreds of emission lines from multiple molecules is at the same time very promising and still very challenging, as demonstrated by the new spectral forests observed with the MIRI spectrograph \citep{miri,miri2} on the James Webb Space Telescope (JWST) \citep[e.g.][]{grant23,gasman23,banz23b,pontoppidan23}.

iSLAT has been built with this potential and challenge in mind, to provide the community with a tool that enables real-time interaction with infrared molecular spectra by comparison to simulated spectra. To handle the combination of large wavelength range and spectral detail of hundreds of lines provided by modern infrared spectrographs, iSLAT features a fully interactive graphical user interface (GUI) that allows users to zoom in/out and pan across the spectra, adjust model parameters for multiple molecules in real time, compare simulated spectra to the data, and perform a number of analysis actions that will be described below. 

\begin{figure*}
\centering
\includegraphics[width=1\textwidth]{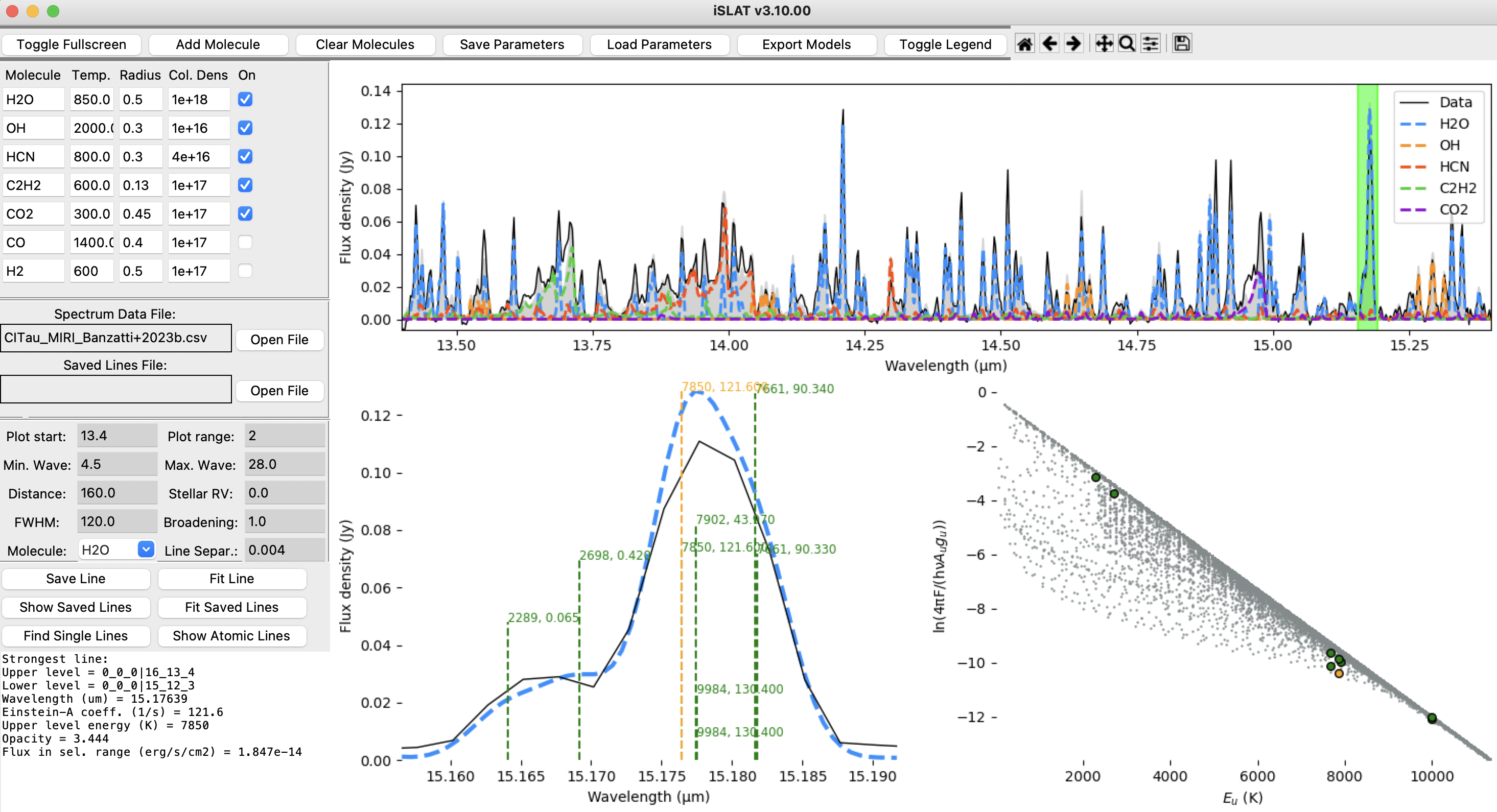}
\caption{Screenshot of iSLAT's GUI, using as an example the JWST-MIRI spectrum of CI~Tau from \cite{banz23b}.
\textit{Top graph}: observed spectrum (black line) and overlaid model spectra (colored lines). The gray shaded region is the sum of all model spectra selected for plotting. 
\textit{Bottom left graph}: Zoomed-in view of the data and \ce{H2O} model spectrum in the selected wavelength region (light green region in the spectrum graph above). Individual \ce{H2O} transitions are shown with vertical dashed lines with height proportional to transition intensity for the specific model shown. The strongest line is shown in orange. 
\textit{Bottom right graph}: \ce{H2O} population diagram for the entire model. The specific lines selected in the bottom left are highlighted in the population diagram too.
\textit{Left}: The left side of iSLAT's GUI provides the user with control over all model parameters and functions. The text box reports action messages and the properties of the strongest line in the selected region.}
\label{fig: iSLAT_miri_example}
\end{figure*}

The slab model code for generating the molecular spectra in iSLAT was written in python by Simon Bruderer and described in the Appendix of \citet{banz12}. The model has four fundamental parameters: the excitation temperature $T$ and the column density $N$ that determine line ratios, the slab surface area $A$ with equivalent emitting radius $R$ that scales all lines equally, and a line width from thermal and turbulent broadening. The synthetic spectra are then scaled to the distance of the specific object, convolved with the instrumental resolution, and shifted to the stellar radial velocity (RV). All these parameters are controlled from the GUI.

iSLAT currently uses line parameters from the HITRAN database and can therefore include any molecule offered with the current HITRAN release \citep{hitran20}. A list of six common molecules is loaded into iSLAT by default: CO, \ce{H2O}, \ce{OH}, \ce{HCN}, \ce{C2H2}, and \ce{CO2}. Users can then add any other molecules from the HITRAN database as needed in their analyses. Users may also add multiple instances of the same molecule with their own independent parameters. This may be useful, for example, to explore how each parameter affects the molecule spectral excitation in different lines, or when multiple components (or a temperature gradient) of the same molecule are present \citep[e.g.][]{banz23,banz23b,gasman23}. 

In its conception, iSLAT has been designed to support the analysis of molecular spectra obtained with space telescopes (Spitzer and JWST) and ground-based high-resolution spectrographs (CRIRES, iSHELL, NIRSPEC), with specific reference to the dense forests of lines observed from protoplanetary disks \citep[see e.g. the case of water spectra as observed with multiple instruments in][]{banz23}. iSLAT's flexible structure, however, makes it a versatile tool for molecular spectra observed with other past, present, and future instruments.

\section{iSLAT's GUI and its features} \label{sec: data}
iSLAT's GUI is illustrated in Figure \ref{fig: iSLAT_miri_example}. At the top of the GUI, an interactive {\tt matplotlib} graph is used to visualize spectra that the user can upload in .csv file format. The input spectrum is assumed to be continuum-subtracted (this needs to be done before using iSLAT\footnote{E.g. using {\tt https://github.com/pontoppi/ctool} presented in \citet{pontoppidan23}}) and to have a wavelength (``wave", in $\mu$m) and flux (``flux", in Jy) arrays. The input spectrum is plotted on the interactive graph that allows users to pan and zoom on any wavelength region of the spectrum. Within the same graph, iSLAT generates simulated model spectra for multiple molecules that can be overlaid on the data. All model parameters including the temperature (T), column density (N), and equivalent emission radius (R), can be controlled from the GUI for each molecule independently. Any changes in these parameters will adjust the model spectra in the graph in real time. The observed resolution and flux of model spectra is controlled by other parameters in the GUI, including the distance to the observed object and the full width at half maximum (FWHM) of the observed lines, whether that is set by the gas kinematics (if observed at high resolution from the ground) or by the resolving power of the instrument (as in the case of Spitzer and JWST). The model parameters (T, N, R) set by the user for multiple molecules can be saved and loaded for any specific input data spectrum (using the file name as identifier) to allow users to quickly start where they left off at each startup of iSLAT. A text box at the bottom left of the GUI reports messages when specific actions are performed in the GUI (e.g. the temperature of a molecule is updated) and information on any lines selected in the top graph. Additional information from specific functions is printed on terminal, as explained below.

\subsection{Spectral-line inspection}
Users can select any range in the spectrum graph for inspection at high resolution by clicking and dragging a region (light green region of the top graph in Figure \ref{fig: iSLAT_miri_example}). This is one of the central functionalities that motivated the design of iSLAT due to its importance to study molecular spectra as observed at moderate or low resolving powers, as in the case of Spitzer-IRS and JWST-MIRI, where many of the observed emission lines are blend of transitions from multiple energy levels or even from different molecules. After defining a region in the top graph, a zoomed-in version of the data and model spectra in that region will be visualized in the lower-left graph in the GUI. This new graph will visualize the individual transitions as observed at infinite resolution, each one labeled with its upper-level energy and Einstein-$A$ coefficient. The height of transitions visualized in this graph is proportional to the line intensity, with the strongest line in the selected region used as reference. For example, a line that has half the intensity of the strongest line will be half as tall in this graph. Lines that are weaker than 1\% of the strongest line are not plotted, to avoid over-crowding the plot.

The properties of the strongest line will be printed in the text box to the left of the graph, including the line upper and lower level quantum numbers, the upper-level energy, the transition wavelength, Einstein-$A$ coefficient, and the opacity at line center \citep[for its definition, see the appendix in][]{banz12}. A measurement of the line flux is also provided here, as measured by integration of the area below the observed pixels within the region defined by the user. This allows users, for example, to save the flux value for specific regions of interest and use these measurements in their analyses. The properties of the other lines in the selected range are printed on terminal to allow user inspection of weaker lines. The molecule considered in this function is controlled by the drow-down menu in the GUI, i.e. if a user wishes to analyze the spectrum of a molecule other than water (the default), this molecule should be selected in the menu.

\begin{figure}
\centering
\includegraphics[width=0.45\textwidth]{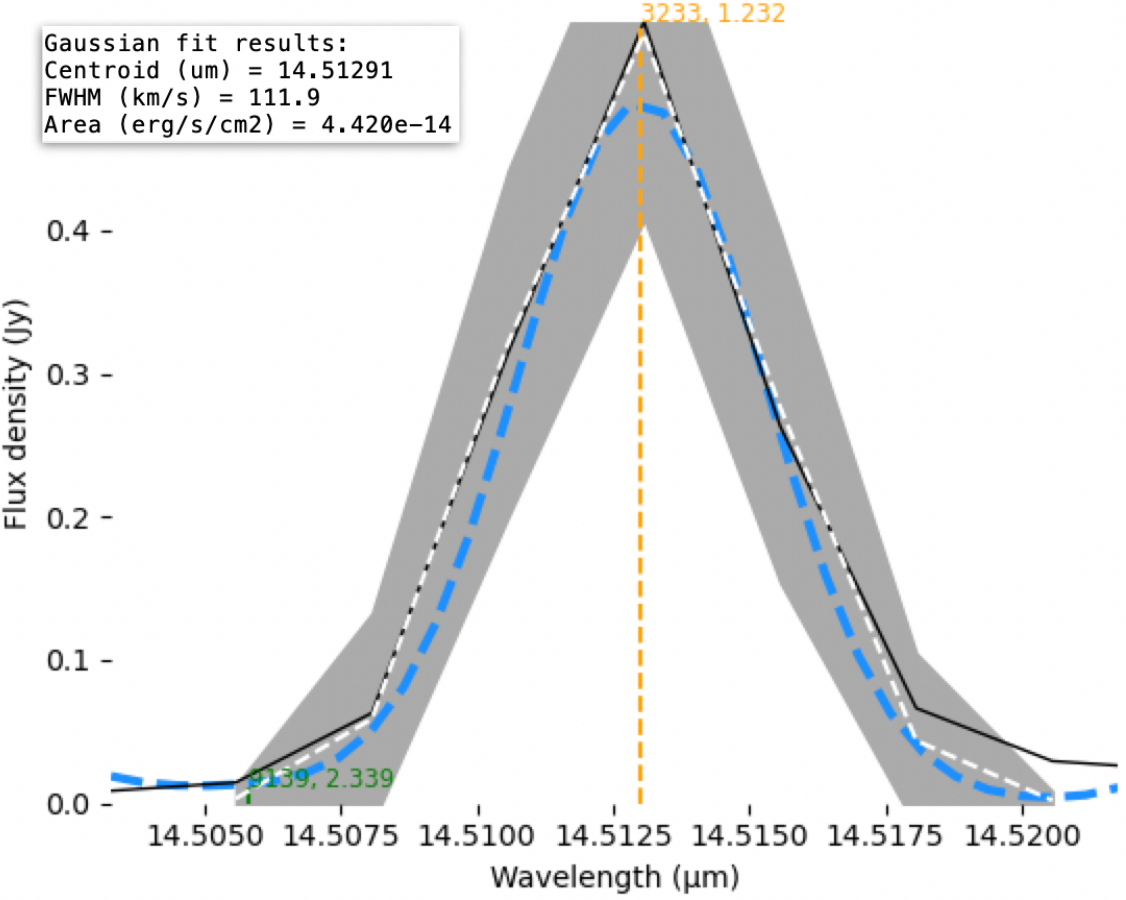}
\caption{Example of a single line fit using the ``Fit Line" function in the GUI. The fit is performed using {\tt lmfit} and currently adopts a single Gaussian function, which is overlaid as a white dashed line on top of the black data line (the grey area shows the fit uncertainty). The fit results reported in iSLAT's message box are included in the inset. }
\label{fig: line_fit}
\end{figure}

The line selected as described above can also be fitted with a Gaussian function using the ``Fit Line" button; the fit is done using {\tt lmfit} and the fit results are added to the message box (Figure \ref{fig: line_fit}). The full fit report from {\tt lmfit}, which includes the uncertainties on model parameters, is printed on terminal for reference. The line centroid from the fit can be used to check for any Doppler shift of the observed lines, for instance due to stellar RV or gas outflows. The line FWHM of individual unresolved lines can be used as a measurement of the instrumental resolving power, e.g. $\sim 110$~km/s in the example shown in Figure \ref{fig: line_fit}. This method has been applied to estimate the MIRI resolving power from observed water lines in \citet{pontoppidan23}.

\begin{figure*}
\centering
\includegraphics[width=0.95\textwidth]{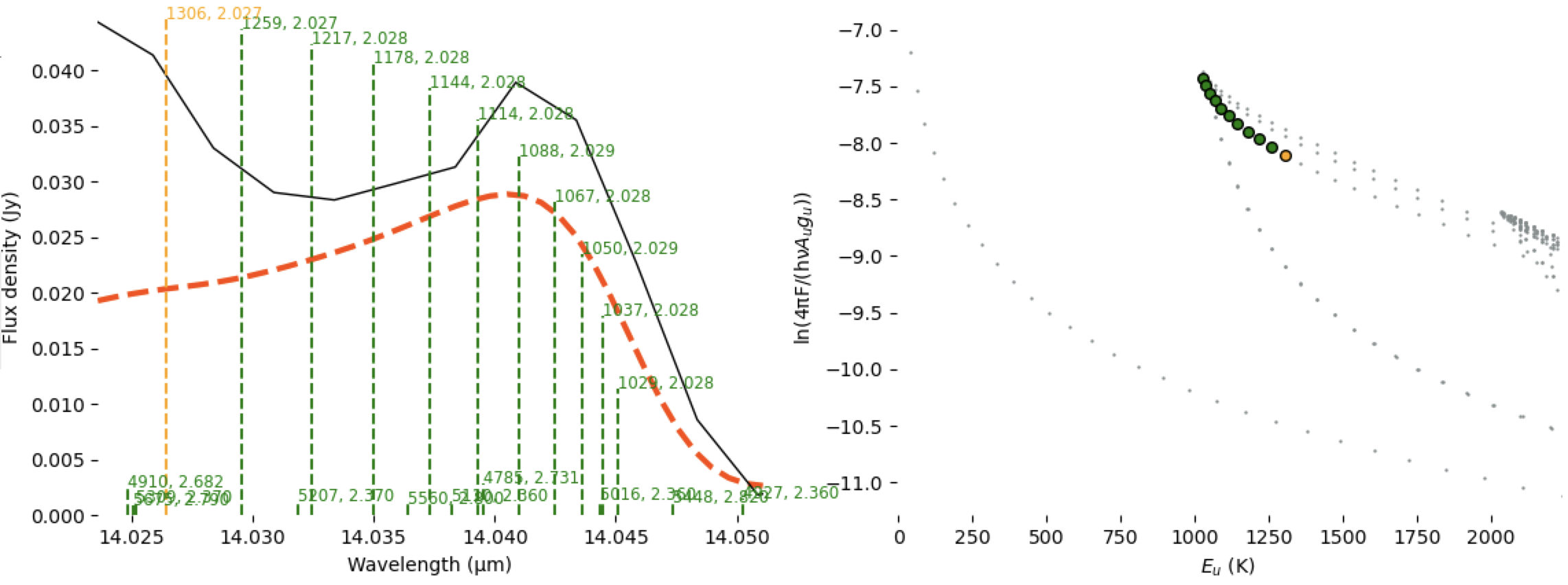}
\caption{Example of inspection of ro-vibrational lines in the $Q$-branch of HCN. These lines can be blended to water and OH lines, when emission from these molecules is present.}
\label{fig: HCN_example}
\end{figure*}

The strongest transition in a chosen spectral region selected as described above can then be appended to a .csv file using the ``Save Line" function in the GUI. To use this function, the user has first to select or define an output file under ``Saved Lines File". The line properties will then be appended to the selected .csv file. Any .csv file of previously saved lines can be loaded into iSLAT allowing for the user to identify these lines in the spectrum graph by using the ``Show Saved Lines" button. iSLAT's release includes some line lists that are useful for the analysis of water spectra as observed with JWST-MIRI, see Appendix \ref{app: line lists}.

\subsection{Population diagram}
In the lower-right part of the GUI, iSLAT includes a plot of the population diagram for a single molecule at a time, as selected from the drop-down menu. The population diagram (also called rotation diagram) is generated in real time for any model parameters set by the user in the GUI, and visualizes the whole spectrum for inspection of excitation conditions and their effects on the population of lines at different upper level energies \citep[for an overview of the technique, see][]{gl99}. When a user selects a spectral region as described above, all transitions in that region are highlighted in the population diagram too, with the strongest one marked in orange. Figure \ref{fig: HCN_example} shows the example of ro-vibrational lines in the $Q$-branch of HCN.

\subsection{Identification of isolated lines}
iSLAT is built to maximize the prompt and flexible interaction with the data. To this end, the slab models need to run quickly and adjust in real time to user input. The slab model code currently used therefore does not consider the opacity overlap of nearby lines when producing the spectrum. This greatly decreases the computational requirements of the tool and reduces the processing time after adjusting model parameters. However, line opacity overlap has been found to become important for some molecules as observed in protoplanetary disks \citep{tabone23}. 

To address that, iSLAT includes a filter tool to automatically identify isolated lines that do not overlap. These lines and their associated fluxes can then be used as reference points when fitting a model to the data, while overlapping lines will be overestimated by the model when the column density is high enough for those lines to become partially or fully optically thick. The single-line filter works by first grouping lines in the selected spectral region that are within a given wavelength range from each other (set by the ``Line Separation" parameter). Lines that are alone in their group are considered isolated and the filter highlights them in the spectrum graph. For all other groups, the strongest line is identified. The filter then considers the strengths of the other lines in the same group. If the other lines have intensities below a desired percentage (currently set to 10\%) of the strongest line in the group, then the strongest line is still considered to be isolated and is highlighted in the spectrum graph. This line selection functionality can be useful for a number of applications besides the line opacity overlap, for instance when users wish to use specific lines or line ratios at different energies as proxies for excitation conditions (e.g. temperature or column density) in modeling explorations.

\begin{figure*}
\centering
\includegraphics[width=1\textwidth]{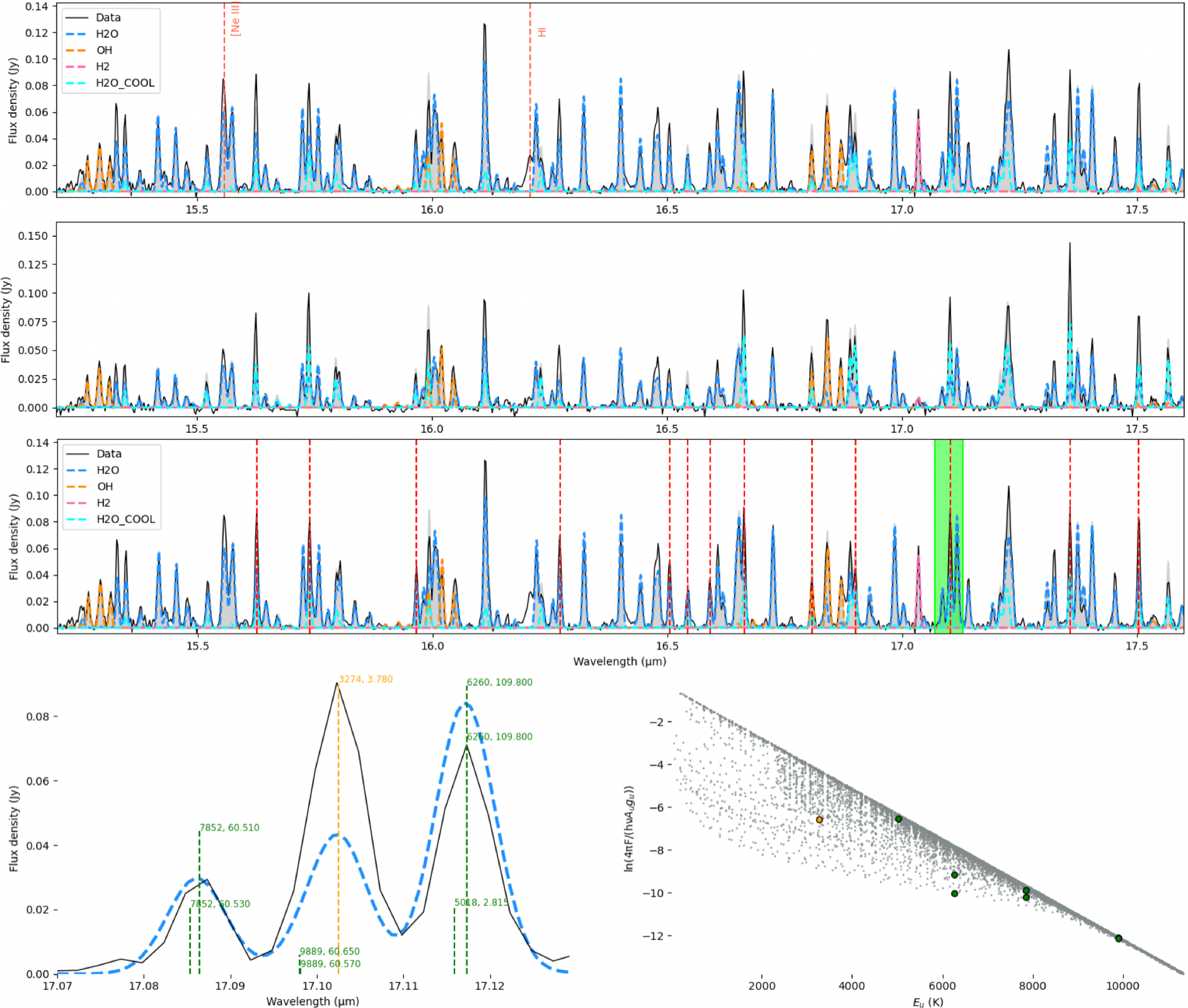}
\caption{Application of iSLAT to study water temperature components in protoplanetary disks as done in \citet{banz23b}. At the top, the hotter-water dominated case of the large, multi-gapped disk of CI~Tau. At the center, the cooler-water dominated case of the compact disk of GK~Tau. The two temperatures used for water in this figure are 850~K (in light blue) and 400~K (in cyan, labeled as ``H2O cool"). In the third plot, an example of using the ``Find Single Lines" function illustrates how even emission lines that look ``single" at the resolution of MIRI can be blends of multiple transitions. Only the central line in the selected example is isolated enough to be measured as a single line flux. The model for the hot water emission under-predicts this line, which requires cooler water from a larger emitting area \citep[for more details, see the analysis in][]{banz23b}.}
\label{fig: water_application}
\end{figure*}

\section{Example Applications}
In this section, we present two example applications of iSLAT to analyze molecular spectra observed from space and from the ground. 

\subsection{Water emission and line blends in MIRI spectra} \label{sec: miri_example}
A first important application of iSLAT is the analysis of the complex spectrum of water as observed with JWST-MIRI. Hundreds of rotational lines are spread across infrared wavelengths, but most of them are close enough to be spectrally blended with lines from other levels when observed at the resolution of MIRI. Figure \ref{fig: water_application} shows the case of two protoplanetary disks where the water spectrum is dominated by hotter (850~K) vs colder (400~K) water emission, the disks of CI~Tau and GK~Tau from \citet{banz23b}. With iSLAT, it is easy to compare their spectra to models of water at different temperatures to check which thermal component dominates the observed flux in different emission lines. Blending with other molecules (the case of OH transitions is shown in the figure) can also easily be checked, as well as emission from atomic species that are prominent in some disk spectra (CI~Tau in this example). 

\begin{figure*}
\centering
\includegraphics[width=1\textwidth]{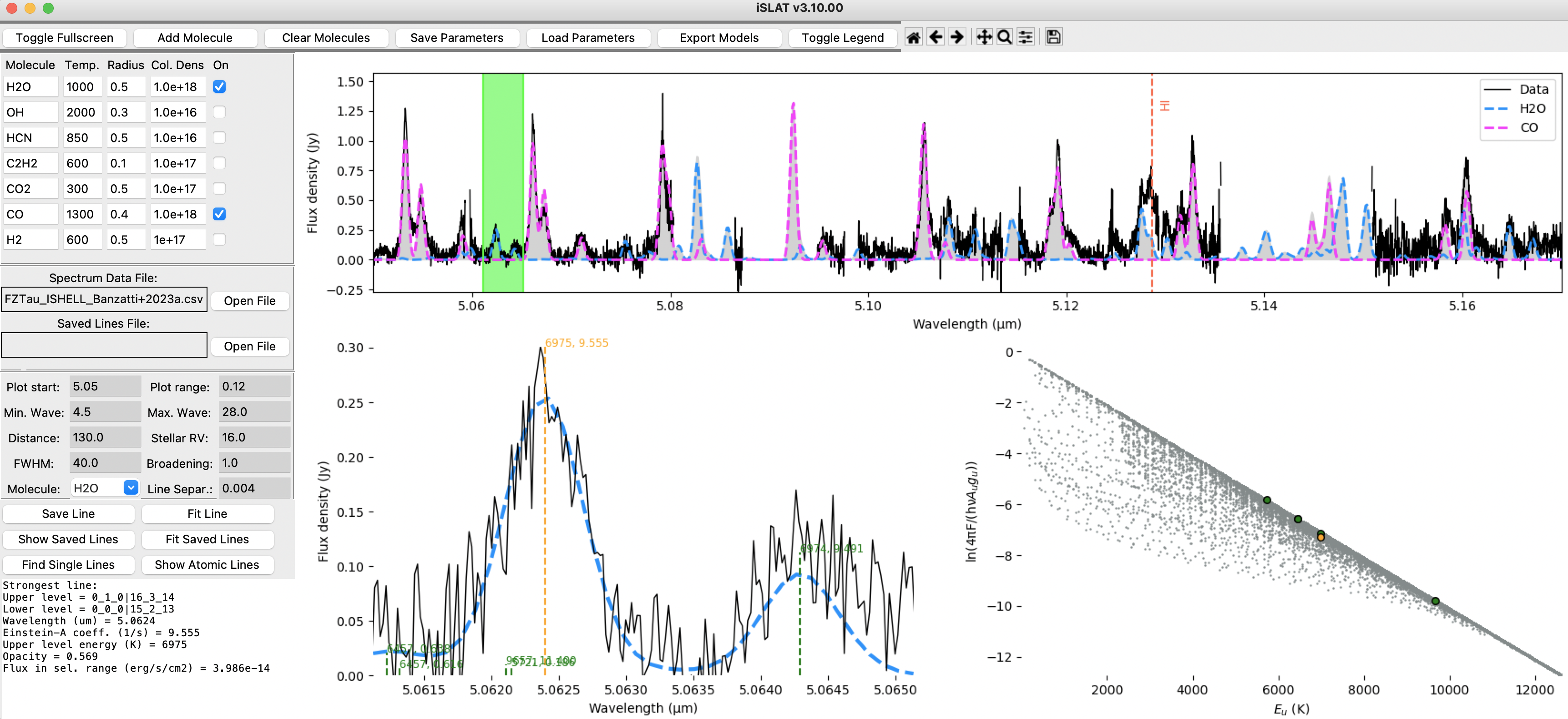}
\caption{Application of iSLAT to a high-resolution $M$-band spectrum from IRTF-iSHELL for the disk of FZ~Tau from \citet{banz23}. A stellar RV of 16 km/s has been applied to shift the data, as measured from optical spectra \citep{banz19}. Water lines at these wavelength are predominantly from the fundamental $v=1-0$ band. The gaps in the observed spectrum are due to detector gaps or low telluric transmission.}
\label{fig: iSLAT_Mband_example}
\end{figure*}

The bottom of Figure \ref{fig: water_application} illustrates an example of inspecting water lines for blending from different energy transitions. Using the ``Find Single Lines" function, transitions that are not significantly blended with transitions from the same molecule but different energy levels will be marked in the spectrum with a vertical dashed line. Inspection of a complex of three nearby lines shows why only one of them is marked as single: the other two are blends of ortho-para pairs of lines, and in one case include contamination from a lower energy level too. Even at the resolution of MIRI, most of the observed water lines are actually blends of multiple transitions, and iSLAT allows users to identify the single ones in one click. Users may want to explore using different line separation limits (the ``Line Separ." parameter in the GUI) to make sure the line selection is appropriate in different wavelength and resolution conditions. For instance, while a line separation of 0.004~$\mu$m seems to be appropriate for water lines in the 10--27~$\mu$m range of MIRI spectra, a line separation of 0.0011--0.0015~$\mu$m seems more appropriate to identify isolated lines in the ro-vibrational band at 5--8~$\mu$m. Appendix \ref{app: line lists} provides useful lists of isolated water lines relevant for JWST-MIRI spectra and used in recent analyses of protoplanetary disks \citep{banz23b}. These line lists are also provided to users in the iSLAT GitHub repository, ready to be loaded and visualized in the GUI.

\subsection{CO and \ce{H2O} spectra at high resolving power}
Figure \ref{fig: iSLAT_Mband_example} shows the application of iSLAT to analyze the CO and \ce{H2O} emission spectra observed from FZ~Tau with the iSHELL spectrograph on IRTF \citep{ishell,ishell22} in \citet{banz23}. In iSLAT, we can now use the FWHM parameter to match the observed line width as broadened by Keplerian rotation in the disk (in this case, about 40 km/s). The other functions demonstrated above can be used in this case too, to identify atomic lines and inspect line blends. One additional parameter that is essential in the case of high-resolution spectra is the stellar radial velocity (RV) in Heliocentric frame, which will shift the observed spectrum to correct for Doppler shift with the emitting source. In the case of FZ~Tau, the RV is about 16 km/s \citep{banz19}. At these wavelengths water transitions belong to ro-vibrational fundamental bands, predominantly from the first vibrational level down to ground ($v=1-0$) but including some from the second level too ($v=2-1$).



\section{Tool release} 
iSLAT is shared with the community on the GitHub page of the SpExoDisks organization ({\tt https://github.com/spexod/iSLAT}). A README file guides users through installation and definitions of parameters and functions. A few example spectra from MIRI, iSHELL, and IRS are included in the tool release for users to practice and get familiar with iSLAT; the spectra are continuum subtracted and were originally published in \citet{pont10, banz20, banz23, banz23b, pontoppidan23}. The original spectra are available on {\tt https://spexodisks.com}.
We invite the scientific community to contribute to improving and expanding iSLAT on GitHub.


\facilities{JWST, IRTF}

\software{
Matplotlib \citep{matplotlib}, NumPy \citep{numpy}, SciPy \citep{scipy}, Astropy \citep{astropy}
}

\appendix

\section{Useful line lists for the analysis of water spectra from JWST-MIRI} \label{app: line lists}
The tables in this Section report selected water line lists that are useful for the analysis of JWST-MIRI spectra; the full tables are provided to users as .csv files that can be loaded into iSLAT as part of the tool release on GitHub. Water is an asymmetric top molecule with three vibrational ($v_1 ~ v_2 ~ v_3$, for symmetric stretching, bending, and asymmetric stretching modes respectively) and three rotational ($J ~ K_a ~ K_c$) quantum numbers, which are included in the table. Transition data are from the current HITRAN release \citep{hitran20}.

Table \ref{tab: single lines} reports single/isolated rotational lines in the ground vibrational level ($v=0-0$) at 10--28~$\mu$m that have been selected using the ``Single Lines" function as described in Section \ref{sec: miri_example}, and additionally filtered to avoid blending with emission from other common molecules.

Table \ref{tab: o-p pairs} reports ortho-para line pairs in rotational transitions at 10--28~$\mu$m. Most of these transitions are close enough to overlap and contribute to the total opacity of the observed spectral line (see examples in Figure \ref{fig: water_application}).

Table \ref{tab: 1-1 lines} reports rotational lines in the first vibrational level ($v=1-1$) emitting at 10--28~$\mu$m. These lines can be excited enough to be detected in protoplanetary disks, but may show sub-excitation due to non-LTE conditions \citep{banz23}. LTE model fits to the spectra may therefore show a discrepancy to the data at these wavelengths, see e.g. the lines at 16.93, 17.31, and 17.68~$\mu$m that are over-predicted by the model in Figure \ref{fig: water_application}. 

Table \ref{tab: single 1-0 lines} reports single/isolated ro-vibrational lines at 5--8.5~$\mu$m that have been selected using the ``Single Lines" function as described in Section \ref{sec: miri_example}. The ro-vibrational bands at these wavelengths are also usually detected in disks, but typically show sub-excitation due to non-LTE conditions \citep{bosman22,banz23}.

\begin{deluxetable}{l l c c}
\tabletypesize{\small}
\tablewidth{0pt}
\tablecaption{\label{tab: single lines} List of strong isolated (single) H$_2$O rotational transitions available at JWST-MIRI wavelengths.}
\tablehead{\colhead{Wavelength} & \colhead{Transition (upper-lower levels)} & \colhead{$A_{ul}$} & \colhead{$E_{u}$} \\
\colhead{($\mu$m)} & \colhead{(level format: $v_1 v_2 v_3~~J_{\:K_a \:K_c}$)} & \colhead{(s$^{-1}$)} & \colhead{(K)}}
\tablecolumns{4}
\startdata
10.1132 & 000-000 \: $17_{\:7\:10} - 16_{\:4\:13}$ & 1.558 & 6371 \\
10.76435 & 000-000 \: $14_{\:9\:6} - 13_{\:6\:7}$ & 0.7628 & 5302 \\
10.85307 & 000-000 \: $15_{\:6\:9} - 14_{\:3\:12}$ & 0.6586 & 4996 \\
11.00168 & 000-000 \: $12_{\:6\:7} - 11_{\:1\:10}$ & 0.06643 & 3501 \\
11.17771 & 000-000 \: $20_{\:8\:13} - 19_{\:5\:14}$ & 14.51 & 8556 \\
11.2531 & 000-000 \: $12_{\:9\:4} - 11_{\:6\:5}$ & 0.2919 & 4363 \\
11.26877 & 000-000 \: $20_{\:7\:14} - 19_{\:4\:15}$ & 16.82 & 8257 \\
11.61657 & 000-000 \: $18_{\:8\:11} - 17_{\:5\:12}$ & 7.511 & 7244 \\
11.64764 & 000-000 \: $17_{\:3\:14} - 16_{\:2\:15}$ & 4.608 & 5483 \\
11.70161 & 000-000 \: $13_{\:5\:8} - 12_{\:2\:11}$ & 0.1994 & 3783 \\
11.96812 & 000-000 \: $14_{\:8\:7} - 13_{\:5\:8}$ & 1.268 & 4985 \\
12.26544 & 000-000 \: $18_{\:7\:12} - 17_{\:4\:13}$ & 12.28 & 6953 \\
12.5645 & 000-000 \: $10_{\:6\:5} - 9_{\:1\:8}$ & 0.03912 & 2697 \\
12.72945 & 000-000 \: $17_{\:7\:11} - 16_{\:4\:12}$ & 9.669 & 6344 \\
12.89409 & 000-000 \: $12_{\:5\:7} - 11_{\:2\:10}$ & 0.2562 & 3310 \\
12.98575 & 000-000 \: $12_{\:7\:5} - 11_{\:4\:8}$ & 0.7422 & 3759 \\
13.13243 & 000-000 \: $16_{\:7\:10} - 15_{\:4\:11}$ & 7.028 & 5763 \\
13.29319 & 000-000 \: $15_{\:3\:12} - 14_{\:2\:13}$ & 3.783 & 4431 \\
13.50312 & 000-000 \: $11_{\:7\:4} - 10_{\:4\:7}$ & 0.4852 & 3340 \\
13.91485 & 000-000 \: $15_{\:5\:11} - 14_{\:2\:12}$ & 6.253 & 4704 \\
14.19856 & 000-000 \: $20_{\:10\:11} - 19_{\:9\:10}$ & 80.31 & 9218 \\
14.34608 & 000-000 \: $14_{\:3\:11} - 13_{\:2\:12}$ & 3.382 & 3941 \\
14.42757 & 000-000 \: $15_{\:4\:11} - 14_{\:3\:12}$ & 6.089 & 4668 \\
14.51301 & 000-000 \: $13_{\:2\:11} - 12_{\:1\:12}$ & 1.232 & 3232 \\
14.60337 & 000-000 \: $19_{\:10\:10} - 18_{\:9\:9}$ & 78.26 & 8547 \\
14.89513 & 000-000 \: $14_{\:5\:10} - 13_{\:2\:11}$ & 5.491 & 4198 \\
15.62568 & 000-000 \: $13_{\:3\:10} - 12_{\:2\:11}$ & 2.988 & 3474 \\
15.73819 & 000-000 \: $12_{\:3\:10} - 11_{\:0\:11}$ & 1.102 & 2823 \\
15.96622 & 000-000 \: $13_{\:5\:9} - 12_{\:2\:10}$ & 4.676 & 3721 \\
16.27136 & 000-000 \: $15_{\:5\:10} - 14_{\:4\:11}$ & 9.232 & 4835 \\
16.50525 & 000-000 \: $17_{\:7\:10} - 16_{\:6\:11}$ & 28.79 & 6371 \\
16.54402 & 000-000 \: $11_{\:6\:6} - 10_{\:3\:7}$ & 1.37 & 3082 \\
16.59123 & 000-000 \: $16_{\:9\:7} - 15_{\:8\:8}$ & 56.44 & 6369 \\
16.66379 & 000-000 \: $12_{\:4\:9} - 11_{\:1\:10}$ & 2.691 & 3057 \\
16.8082 & 000-000 \: $8_{\:6\:2} - 7_{\:3\:5}$ & 0.1388 & 2031 \\
16.90054 & 000-000 \: $9_{\:5\:4} - 8_{\:2\:7}$ & 0.2601 & 2125 \\
17.10254 & 000-000 \: $12_{\:5\:8} - 11_{\:2\:9}$ & 3.78 & 3273 \\
17.35766 & 000-000 \: $11_{\:2\:9} - 10_{\:1\:10}$ & 0.9617 & 2432 \\
17.50436 & 000-000 \: $13_{\:4\:9} - 12_{\:3\:10}$ & 4.941 & 3645 \\
18.25429 & 000-000 \: $11_{\:5\:7} - 10_{\:2\:8}$ & 2.805 & 2857 \\
18.33865 & 000-000 \: $8_{\:4\:4} - 7_{\:1\:7}$ & 0.07159 & 1628 \\
19.12996 & 000-000 \: $15_{\:7\:9} - 14_{\:6\:8}$ & 27.0 & 5214 \\
19.24597 & 000-000 \: $11_{\:3\:8} - 10_{\:2\:9}$ & 2.275 & 2608 \\
19.34995 & 000-000 \: $10_{\:5\:6} - 9_{\:2\:7}$ & 1.829 & 2472 \\
19.68805 & 000-000 \: $14_{\:7\:8} - 13_{\:6\:7}$ & 27.3 & 4696 \\
\enddata
\tablecomments{Line properties are from HITRAN \citep{hitran20}. The full table is available on GitHub.}
\end{deluxetable}

\begin{deluxetable}{l l c c}
\tabletypesize{\small}
\tablewidth{0pt}
\tablecaption{\label{tab: o-p pairs} List of strong H$_2$O ortho-para line pairs available at JWST-MIRI wavelengths.}
\tablehead{\colhead{Wavelength} & \colhead{Transition (upper-lower levels)} & \colhead{$A_{ul}$} & \colhead{$E_{u}$} \\
\colhead{($\mu$m)} & \colhead{(level format: $v_1 v_2 v_3~~J _{\:K_a \: K_c}$)} & \colhead{(s$^{-1}$)} & \colhead{(K)}}
\tablecolumns{4}
\startdata
11.03377 & 000-000 \: $17_{\:3\:15} - 16_{\:0\:16}$ & 1.85 & 5132 \\
11.03475 & 000-000 \: $17_{\:2\:15} - 16_{\:1\:16}$ & 1.85 & 5132 \\
11.72455 & 000-000 \: $16_{\:3\:14} - 15_{\:0\:15}$ & 1.682 & 4620 \\
11.72672 & 000-000 \: $16_{\:2\:14} - 15_{\:1\:15}$ & 1.682 & 4619 \\
13.55068 & 000-000 \: $20_{\:12\:8} - 19_{\:11\:9}$ & 118.9 & 9978 \\
13.55091 & 000-000 \: $20_{\:12\:9} - 19_{\:11\:8}$ & 118.9 & 9978 \\
13.55404 & 000-000 \: $19_{\:14\:6} - 18_{\:13\:5}$ & 154.2 & 10121 \\
13.55404 & 000-000 \: $19_{\:14\:5} - 18_{\:13\:6}$ & 154.1 & 10121 \\
13.73089 & 000-000 \: $19_{\:13\:6} - 18_{\:12\:7}$ & 134.3 & 9708 \\
13.7309 & 000-000 \: $19_{\:13\:7} - 18_{\:12\:6}$ & 134.4 & 9708 \\
13.76219 & 000-000 \: $18_{\:16\:2} - 17_{\:15\:3}$ & 190.2 & 10297 \\
13.76219 & 000-000 \: $18_{\:16\:3} - 17_{\:15\:2}$ & 190.2 & 10297 \\
13.8552 & 000-000 \: $18_{\:15\:3} - 17_{\:14\:4}$ & 169.5 & 9882 \\
13.8552 & 000-000 \: $18_{\:15\:4} - 17_{\:14\:3}$ & 169.4 & 9882 \\
13.95566 & 000-000 \: $19_{\:12\:7} - 18_{\:11\:8}$ & 115.1 & 9305 \\
13.9557 & 000-000 \: $19_{\:12\:8} - 18_{\:11\:7}$ & 115.1 & 9305 \\
13.98964 & 000-000 \: $18_{\:14\:4} - 17_{\:13\:5}$ & 149.4 & 9468 \\
13.98964 & 000-000 \: $18_{\:14\:5} - 17_{\:13\:4}$ & 149.4 & 9468 \\
14.16883 & 000-000 \: $18_{\:13\:5} - 17_{\:12\:6}$ & 130.0 & 9059 \\
14.16883 & 000-000 \: $18_{\:13\:6} - 17_{\:12\:5}$ & 130.0 & 9059 \\
14.21063 & 000-000 \: $17_{\:17\:1} - 16_{\:16\:0}$ & 205.7 & 10062 \\
14.21063 & 000-000 \: $17_{\:17\:0} - 16_{\:16\:1}$ & 205.6 & 10062 \\
14.23737 & 000-000 \: $19_{\:11\:8} - 18_{\:10\:9}$ & 96.63 & 8916 \\
14.2388 & 000-000 \: $19_{\:11\:9} - 18_{\:10\:8}$ & 96.57 & 8916 \\
14.24955 & 000-000 \: $17_{\:16\:2} - 16_{\:15\:1}$ & 184.7 & 9660 \\
14.24955 & 000-000 \: $17_{\:16\:1} - 16_{\:15\:2}$ & 184.8 & 9660 \\
14.33547 & 000-000 \: $17_{\:15\:3} - 16_{\:14\:2}$ & 164.4 & 9252 \\
14.33547 & 000-000 \: $17_{\:15\:2} - 16_{\:14\:3}$ & 164.4 & 9252 \\
14.39851 & 000-000 \: $18_{\:12\:6} - 17_{\:11\:7}$ & 111.4 & 8660 \\
14.39857 & 000-000 \: $18_{\:12\:7} - 17_{\:11\:6}$ & 111.4 & 8660 \\
14.46772 & 000-000 \: $17_{\:14\:3} - 16_{\:13\:4}$ & 144.7 & 8844 \\
14.46772 & 000-000 \: $17_{\:14\:4} - 16_{\:13\:3}$ & 144.7 & 8844 \\
14.64871 & 000-000 \: $17_{\:13\:5} - 16_{\:12\:4}$ & 125.8 & 8440 \\
14.64871 & 000-000 \: $17_{\:13\:4} - 16_{\:12\:5}$ & 125.8 & 8440 \\
14.68765 & 000-000 \: $18_{\:11\:7} - 17_{\:10\:8}$ & 93.55 & 8274 \\
14.68822 & 000-000 \: $18_{\:11\:8} - 17_{\:10\:7}$ & 93.53 & 8274 \\
14.79081 & 000-000 \: $16_{\:16\:1} - 15_{\:15\:0}$ & 179.5 & 9050 \\
14.79081 & 000-000 \: $16_{\:16\:0} - 15_{\:15\:1}$ & 179.5 & 9050 \\
14.86608 & 000-000 \: $16_{\:15\:1} - 15_{\:14\:2}$ & 159.5 & 8650 \\
14.86608 & 000-000 \: $16_{\:15\:2} - 15_{\:14\:1}$ & 159.5 & 8650 \\
14.88384 & 000-000 \: $17_{\:12\:6} - 16_{\:11\:5}$ & 107.6 & 8044 \\
14.88385 & 000-000 \: $17_{\:12\:5} - 16_{\:11\:6}$ & 107.6 & 8044 \\
14.99421 & 000-000 \: $16_{\:14\:3} - 15_{\:13\:2}$ & 140.1 & 8248 \\
14.99421 & 000-000 \: $16_{\:14\:2} - 15_{\:13\:3}$ & 140.1 & 8248 \\
\enddata
\tablecomments{Line properties are from HITRAN \citep{hitran20}. The full table is available on GitHub.}
\end{deluxetable} 

\begin{deluxetable}{l l c c}
\tabletypesize{\small}
\tablewidth{0pt}
\tablecaption{\label{tab: 1-1 lines} List of strong $v=1-1$ H$_2$O transitions available at JWST-MIRI wavelengths.}
\tablehead{\colhead{Wavelength} & \colhead{Transition (upper-lower levels)} & \colhead{$A_{ul}$} & \colhead{$E_{u}$} \\
\colhead{($\mu$m)} & \colhead{(level format: $v_1 v_2 v_3~~J _{\:K_a \: K_c}$)} & \colhead{(s$^{-1}$)} & \colhead{(K)}}
\tablecolumns{4}
\startdata
10.33669 & 010-010 \: $17_{\:2\:15} - 16_{\:1\:16}$ & 2.598 & 7488 \\
10.43161 & 010-010 \: $18_{\:4\:15} - 17_{\:1\:16}$ & 6.625 & 8433 \\
10.9884 & 010-010 \: $16_{\:3\:14} - 15_{\:0\:15}$ & 2.322 & 6974 \\
11.06768 & 010-010 \: $18_{\:5\:14} - 17_{\:2\:15}$ & 10.52 & 8788 \\
11.08288 & 010-010 \: $17_{\:3\:14} - 16_{\:2\:15}$ & 5.979 & 7864 \\
11.45486 & 010-010 \: $13_{\:7\:6} - 12_{\:4\:9}$ & 1.107 & 6681 \\
11.57343 & 010-010 \: $18_{\:6\:13} - 17_{\:3\:14}$ & 13.54 & 9108 \\
11.73657 & 010-010 \: $15_{\:3\:13} - 14_{\:0\:14}$ & 2.073 & 6485 \\
11.74744 & 010-010 \: $15_{\:2\:13} - 14_{\:1\:14}$ & 2.07 & 6484 \\
11.75123 & 010-010 \: $16_{\:4\:13} - 15_{\:1\:14}$ & 5.417 & 7329 \\
11.76071 & 010-010 \: $13_{\:6\:7} - 12_{\:3\:10}$ & 1.036 & 6392 \\
11.81779 & 010-010 \: $16_{\:3\:13} - 15_{\:2\:14}$ & 5.376 & 7322 \\
12.00584 & 010-010 \: $17_{\:4\:13} - 16_{\:3\:14}$ & 9.3 & 8173 \\
12.10458 & 010-010 \: $16_{\:7\:10} - 15_{\:4\:11}$ & 6.272 & 8230 \\
12.28718 & 010-010 \: $11_{\:7\:4} - 10_{\:4\:7}$ & 0.4983 & 5810 \\
12.40411 & 010-010 \: $14_{\:7\:8} - 13_{\:4\:9}$ & 2.493 & 7165 \\
12.44545 & 010-010 \: $16_{\:5\:12} - 15_{\:2\:13}$ & 8.52 & 7640 \\
12.555 & 010-010 \: $15_{\:4\:12} - 14_{\:1\:13}$ & 4.864 & 6814 \\
12.59914 & 010-010 \: $14_{\:3\:12} - 13_{\:0\:13}$ & 1.845 & 6021 \\
12.60313 & 010-010 \: $12_{\:7\:6} - 11_{\:4\:7}$ & 0.8958 & 6228 \\
12.62215 & 010-010 \: $14_{\:2\:12} - 13_{\:1\:13}$ & 1.84 & 6018 \\
12.68409 & 010-010 \: $15_{\:3\:12} - 14_{\:2\:13}$ & 4.796 & 6802 \\
12.75847 & 010-010 \: $16_{\:6\:11} - 15_{\:3\:12}$ & 9.816 & 7930 \\
12.94375 & 010-010 \: $16_{\:4\:12} - 15_{\:3\:13}$ & 8.302 & 7596 \\
13.15066 & 010-010 \: $11_{\:6\:5} - 10_{\:3\:8}$ & 0.6687 & 5515 \\
13.2367 & 010-010 \: $11_{\:5\:6} - 10_{\:2\:9}$ & 0.4325 & 5265 \\
13.2474 & 010-010 \: $15_{\:5\:11} - 14_{\:2\:12}$ & 7.513 & 7105 \\
13.33791 & 010-010 \: $15_{\:6\:10} - 14_{\:3\:11}$ & 7.738 & 7384 \\
13.47704 & 010-010 \: $14_{\:4\:11} - 13_{\:1\:12}$ & 4.337 & 6324 \\
13.4916 & 010-010 \: $17_{\:5\:12} - 16_{\:4\:13}$ & 12.73 & 8395 \\
13.60264 & 010-010 \: $13_{\:3\:11} - 12_{\:0\:12}$ & 1.636 & 5582 \\
13.65169 & 010-010 \: $13_{\:2\:11} - 12_{\:1\:12}$ & 1.626 & 5578 \\
13.72562 & 010-010 \: $14_{\:3\:11} - 13_{\:2\:12}$ & 4.239 & 6305 \\
13.86514 & 010-010 \: $14_{\:6\:9} - 13_{\:3\:10}$ & 5.628 & 6868 \\
14.08938 & 010-010 \: $15_{\:4\:11} - 14_{\:3\:12}$ & 7.416 & 7042 \\
14.1191 & 010-010 \: $14_{\:5\:10} - 13_{\:2\:11}$ & 6.465 & 6597 \\
14.30909 & 010-010 \: $13_{\:6\:8} - 12_{\:3\:9}$ & 3.717 & 6384 \\
14.33815 & 010-010 \: $17_{\:12\:5} - 16_{\:11\:6}$ & 120.2 & 10707 \\
14.45525 & 010-010 \: $18_{\:10\:9} - 17_{\:9\:8}$ & 87.23 & 10493 \\
14.47554 & 010-010 \: $9_{\:6\:3} - 8_{\:3\:6}$ & 0.2775 & 4778 \\
14.53518 & 010-010 \: $13_{\:4\:10} - 12_{\:1\:11}$ & 3.829 & 5861 \\
14.59253 & 010-010 \: $17_{\:11\:6} - 16_{\:10\:7}$ & 101.9 & 10287 \\
14.63497 & 010-010 \: $16_{\:13\:4} - 15_{\:12\:3}$ & 134.7 & 10546 \\
14.65748 & 010-010 \: $12_{\:6\:7} - 11_{\:3\:8}$ & 2.228 & 5932 \\
14.72218 & 010-010 \: $16_{\:5\:11} - 15_{\:4\:12}$ & 12.17 & 7791 \\
14.78088 & 010-010 \: $12_{\:3\:10} - 11_{\:0\:11}$ & 1.443 & 5168 \\
\enddata
\tablecomments{Line properties are from HITRAN \citep{hitran20}. The full table is available on GitHub.}
\end{deluxetable} 

\begin{deluxetable}{l l c c}
\tabletypesize{\small}
\tablewidth{0pt}
\tablecaption{\label{tab: single 1-0 lines} List of strong ro-vibrational H$_2$O transitions available at JWST-MIRI wavelengths.}
\tablehead{\colhead{Wavelength} & \colhead{Transition (upper-lower levels)} & \colhead{$A_{ul}$} & \colhead{$E_{u}$} \\
\colhead{($\mu$m)} & \colhead{(level format: $v_1 v_2 v_3~~J _{\:K_a \: K_c}$)} & \colhead{(s$^{-1}$)} & \colhead{(K)}}
\tablecolumns{4}
\startdata
5.2357 & 010-000 \: $7_{\:3\:4} - 6_{\:2\:5}$ & 1.27 & 3543 \\
5.34529 & 010-000 \: $10_{\:3\:8} - 9_{\:2\:7}$ & 5.813 & 4420 \\
5.58362 & 010-000 \: $8_{\:2\:7} - 7_{\:1\:6}$ & 7.856 & 3589 \\
5.61179 & 010-000 \: $9_{\:3\:6} - 8_{\:4\:5}$ & 2.009 & 4179 \\
5.63179 & 010-000 \: $7_{\:2\:6} - 6_{\:1\:5}$ & 6.867 & 3335 \\
5.64107 & 010-000 \: $3_{\:3\:0} - 2_{\:2\:1}$ & 5.815 & 2744 \\
5.70511 & 010-000 \: $7_{\:4\:4} - 7_{\:3\:5}$ & 3.115 & 3696 \\
5.80878 & 010-000 \: $8_{\:4\:4} - 8_{\:3\:5}$ & 4.529 & 3987 \\
5.96224 & 020-010 \: $6_{\:0\:6} - 5_{\:1\:5}$ & 20.56 & 5179 \\
6.0059 & 020-010 \: $5_{\:1\:5} - 4_{\:0\:4}$ & 18.77 & 5010 \\
6.01912 & 010-000 \: $6_{\:3\:4} - 5_{\:4\:1}$ & 0.6975 & 3268 \\
6.07016 & 010-000 \: $4_{\:2\:2} - 4_{\:1\:3}$ & 7.676 & 2766 \\
6.07545 & 010-000 \: $3_{\:2\:1} - 3_{\:1\:2}$ & 6.863 & 2617 \\
6.08018 & 020-010 \: $5_{\:1\:4} - 5_{\:0\:5}$ & 6.438 & 5129 \\
6.14316 & 010-000 \: $2_{\:0\:2} - 1_{\:1\:1}$ & 3.778 & 2395 \\
6.17528 & 020-010 \: $2_{\:1\:2} - 1_{\:0\:1}$ & 13.86 & 4658 \\
6.1854 & 010-000 \: $1_{\:1\:0} - 1_{\:0\:1}$ & 10.94 & 2360 \\
6.19356 & 030-020 \: $5_{\:0\:5} - 4_{\:1\:4}$ & 25.99 & 7188 \\
6.20278 & 020-010 \: $3_{\:0\:3} - 2_{\:1\:2}$ & 11.85 & 4732 \\
6.34443 & 010-000 \: $1_{\:0\:1} - 1_{\:1\:0}$ & 12.63 & 2328 \\
6.37028 & 010-000 \: $2_{\:0\:2} - 2_{\:1\:1}$ & 10.99 & 2395 \\
6.37373 & 010-000 \: $2_{\:2\:1} - 3_{\:1\:2}$ & 1.14 & 2506 \\
6.42663 & 010-000 \: $6_{\:4\:3} - 7_{\:3\:4}$ & 0.5998 & 3450 \\
6.43355 & 010-000 \: $5_{\:1\:4} - 5_{\:2\:3}$ & 11.52 & 2878 \\
6.49224 & 010-000 \: $2_{\:1\:2} - 3_{\:0\:3}$ & 7.211 & 2412 \\
6.52896 & 010-000 \: $7_{\:3\:4} - 7_{\:4\:3}$ & 9.451 & 3543 \\
6.97738 & 010-000 \: $9_{\:0\:9} - 9_{\:1\:8}$ & 4.239 & 3614 \\
6.99328 & 010-000 \: $3_{\:2\:2} - 4_{\:3\:1}$ & 8.566 & 2609 \\
7.14692 & 010-000 \: $4_{\:2\:3} - 5_{\:3\:2}$ & 5.629 & 2745 \\
7.21253 & 010-000 \: $7_{\:2\:5} - 8_{\:3\:6}$ & 4.477 & 3442 \\
7.27924 & 010-000 \: $5_{\:3\:2} - 6_{\:4\:3}$ & 8.014 & 3065 \\
7.30659 & 010-000 \: $5_{\:3\:3} - 6_{\:4\:2}$ & 7.835 & 3059 \\
8.0696 & 010-000 \: $10_{\:5\:6} - 11_{\:6\:5}$ & 5.949 & 4867 \\
\enddata
\tablecomments{Line properties are from HITRAN \citep{hitran20}.}
\end{deluxetable} 

\bibliography{islat_paper}{}
\bibliographystyle{aasjournal}
\pagebreak

\end{document}